\documentclass[journal, 10pt]{IEEEtran}
\pdfoutput=1

\usepackage{times}
\usepackage{amsfonts, amssymb, amsmath,url}
\usepackage{graphicx}
\usepackage{multicol}
\usepackage[center]{caption}

\begin{document}

\title{Tejas Simulator : Validation against Hardware}

\author{Smruti R. Sarangi, Rajshekar Kalayappan, Prathmesh Kallurkar and Seep Goel\\
Computer Science and Engineering \\
Indian Institute of Technology, New Delhi, India \\
e-mail: \{srsarangi, csz128281, csz128280, mcs132582\}@cse.iitd.ac.in
}

\maketitle

\begin{abstract}
In this report we show results that validate the Tejas architectural simulator against
native hardware. We report mean error rates of 11.45\% and 18.77\% for the SPEC2006 and Splash2
benchmark suites respectively. These error rates are competitive and in most cases better than
the numbers reported by other contemporary simulators.
\end{abstract}
\section{Introduction}
\begin{figure*}[!tb]
\begin{center}
\includegraphics[width=0.56\textwidth]{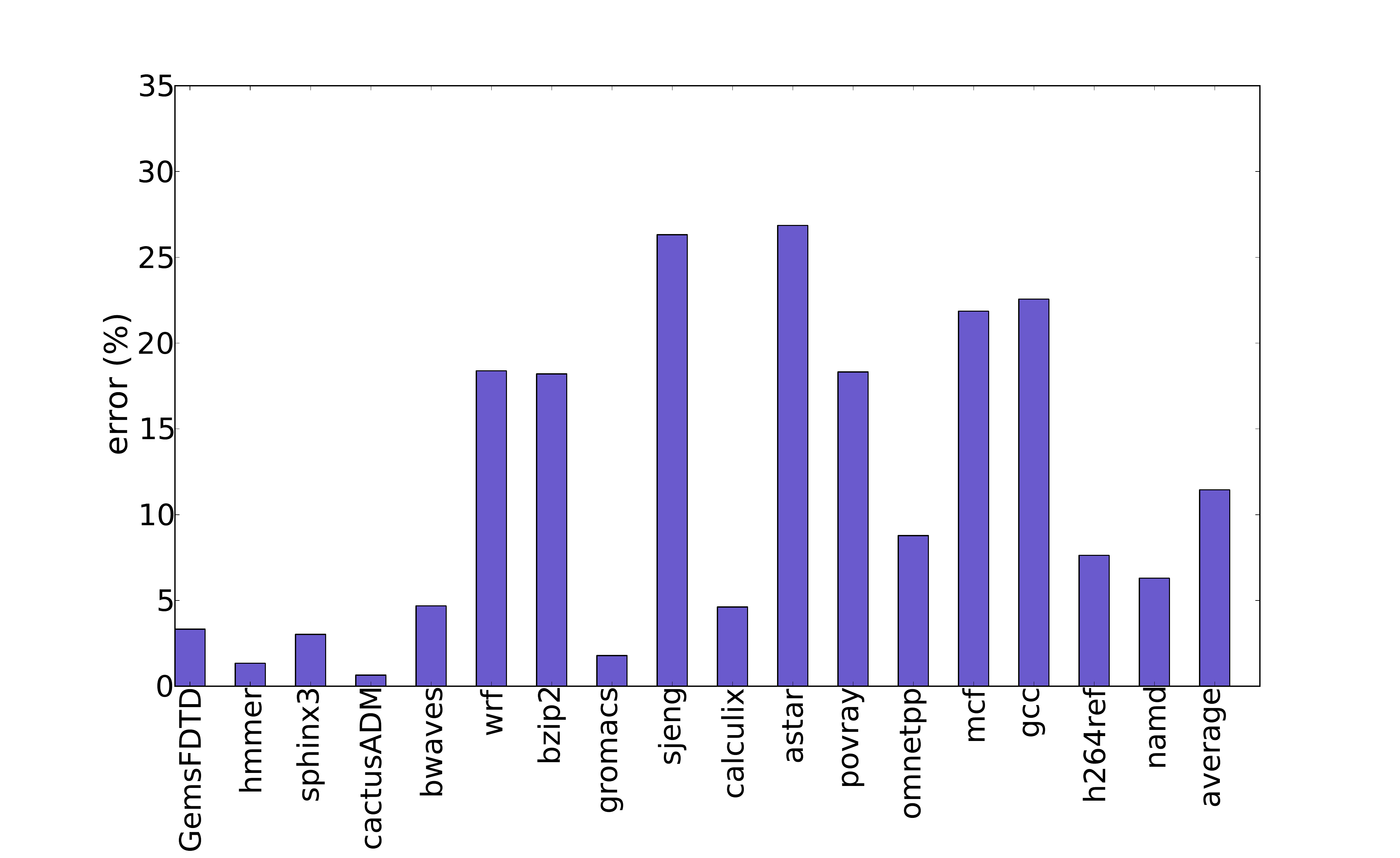}
\caption {Comparison : {\em SPEC2006} suite \label{fig:serial_validation} }
\end{center}
\end{figure*}

\begin{figure*}[!tb]
\begin{center}
\includegraphics[width=0.47\textwidth]{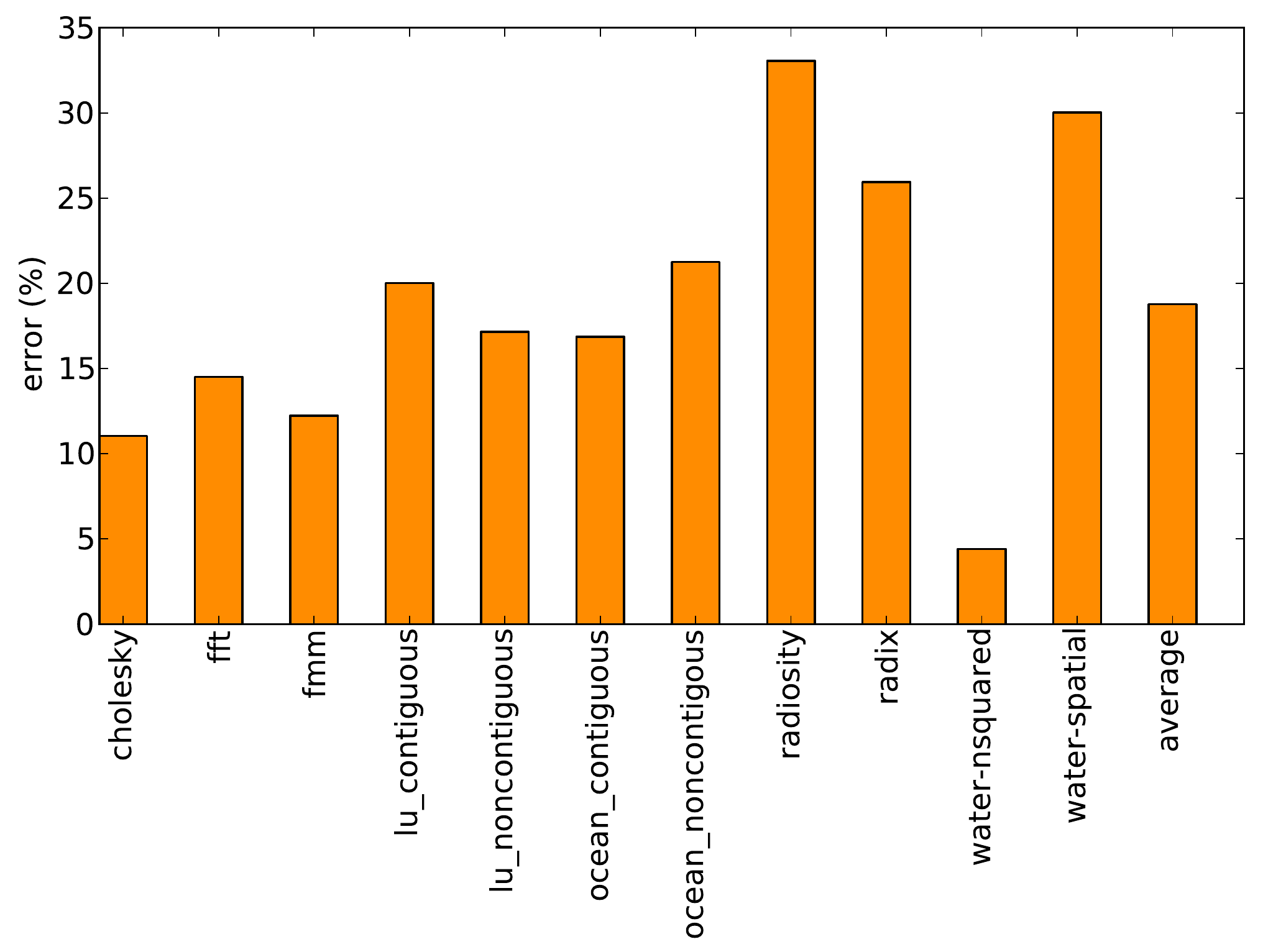}
\caption {Comparison : {\em SPLASH-2} suite \label{fig:parallel_validation} }
\end{center}
\end{figure*}

This article serves to establish Tejas ~\cite{partejas} as a validated micro-architectural simulator.
To do so, a range of serial and parallel benchmarks were run on a PowerEdge R620 server
(for details see Table~\ref{tab:hardware_details}). The linux ``perf'' command was
used to measure the number of cycles taken to execute the benchmarks.
The same set of benchmarks were then run
on the Tejas simulator, configured to mimic the DELL server as closely as possible
(see Table~\ref{tab:tejas_configuration} for details).
The comparison of the simulated cycle counts with their hardware counterparts is given in the next section.

\begin{table}[!htb]
	\centering
	\scriptsize
	\begin{tabular}{|c|c|c|c|}
		\hline
		{\bf Parameter} & {\bf Value} & {\bf Parameter} & {\bf Value} \\
		\hline
		Microarchitecture & Intel Sandybridge & Number of cores & 12 \\
		Main Memory & 32 GB & Memory Type & ECC DDR3 \\
		L1 i-cache and d-cache & 32 KB & L2 cache & 256 KB \\
		L3 Cache & 15 MB & Frequency & 2GHz \\
		Hyper-threading & No & DVFS & Disabled  \\
		Load Buffer Size & 64 & Store Buffer Size & 64  \\
		Reorder Buffer & 168 micro-ops &&\\
\cline{2-4}
		Operating System &  \multicolumn{3}{c|}{ Ubuntu 12.10 Linux 3.5.0-36-generic, 64-bit} \\
		\hline
	\end{tabular}
	\caption{Details of the Reference Hardware \label{tab:hardware_details}}
\end{table}

\section{Results}

Figure~\ref{fig:serial_validation} shows the results for a set of 17 benchmarks from the
SPEC2006 suite. We compute the time it takes for a benchmark to complete on native hardware
(averaged across 10 runs). Then, we simulate the benchmark on Tejas, and compute the absolute error.
The average absolute error is 11.45\%. 10 out of 17 benchmarks have an error less than 10\%.
Only 4 benchmarks have errors in the 20-30\% range ({\em sjeng}, {\em astar}, {\em mcf},
and {\em gcc}).

Figure~\ref{fig:parallel_validation} shows the results for a set of 11 benchmarks from the
{\em SPLASH-2} suite. The average absolute error was observed to be 18.77\%. It has been computed
the same way as was done for the case of sequential benchmarks. In this case, the average
error is more primarily because the jitter introduced by the operating system is not
predictable, there are hardware events that induce jitter, and lastly we are not privy
to all details of the operation of
the cache coherence protocols in Intel systems. Only 3 benchmarks had errors more
than 25\% namely {\em radiosity}, {\em radix} and {\em water-spatial}. For most
of the benchmarks, the error ranges from 10 to 17\%.

\begin{table}
	\scriptsize	
	\centering
	\begin{tabular}{|c|c||c|c|}
		\hline
		{\bf Parameter} & {\bf Value} & {\bf Parameter} & {\bf Value} \\
		\hline
		\multicolumn{4}{|c|}{{\bf Pipeline}} \\
		\hline
		Retire Width & 4 & Integer RF (phy) & 160 \\
		Issue Width & 6 & Float RF (phy) & 144 \\
		ROB size & 168 & Predictor & TAGE~\cite{tage} \\
		IW size & 54 & Bmispred penalty & 8 cycles \\
		LSQ size & 64 & & \\
		iTLB & 128 entries & dTLB & 128 entries \\
		Integer ALU & 3 units & lat = 1 cycle & RoT = 1 \\
		Integer Mul & 1 unit & lat = 3 cycles & RoT = 1 \\
		Integer Div & 1 unit & lat = 21 cycles & RoT = 12\\
		Float ALU & 1 units & lat = 3 cycles & RoT = 1\\
		Float Mul & 1 unit & lat = 5 cycles & RoT = 1 \\
		Float Div & 1 unit & lat = 24 cycles & RoT = 12 \\
	   \hline 
		\multicolumn{4}{|c|}{{\em RoT : reciprocal of throughput}} \\
		\hline
		\hline
		\multicolumn{4}{|c|}{{\bf Private L1 i-cache, d-cache}} \\
		\hline
		Write-mode & Write-Through & Block size & 64 \\
		Associativity & 8 & Size & 32 kB \\
		Latency & 3 cycles & & \\
		\hline
		\hline
		\multicolumn{4}{|c|}{{\bf Private Unified L2 cache}}  \\
		\hline
		Write-mode & Write-Back & Block size & 64 \\
		Associativity & 8 & Size & 256 kB \\
		Latency & 6 cycles & & \\
		\hline
		\multicolumn{4}{|c|}{{\bf Shared L3 cache}}  \\
		\hline
		Write-mode & Write-back & Block size & 64 \\
		Associativity & 8 & Size & 15 MB \\
		Latency & 29 cycles & & \\
		\hline
		\hline
		\multicolumn{2}{|c||}{{\bf Main Memory Latency}} & \multicolumn{2}{c|}{{200 cycles}} \\
		\hline
		\hline
		\multicolumn{4}{|c|}{{\bf NOC and Traffic}}  \\
		\hline
		Topology & Bus & Latency & 1 cycle \\
		Flit size & 32 bytes & &  \\
		\hline
	\end{tabular}
	\caption{Simulation parameters \label{tab:tejas_configuration}}
\end{table}

\section{Comparison with other Simulators}
Let us now put our numbers in the right perspective by comparing similar numbers
obtained on other simulators. We shall observe that Tejas is more accurate on
both serial and parallel benchmarks as compared to most of the other widely
used architecture simulators (for which published results are available).

MARSS~\cite{MARSS} is a cycle-accurate simulator based on PTLSim. 
It is a tool built on QEMU and provides fast and full system simulation.
MARSS has been validated against a x86 target machine
with the Intel Xeon E5620 processor. For the {\em SPEC CPU 2006} benchmark suite,
it has errors ranging from -59.2\% to 50\%,  with an average absolute error of 23.46\%.

Sniper~\cite{sniper} is an approximate simulator.
It tries to find a middle ground between the simulators that are fast but inaccurate
and the simulators that are accurate but slow.
It has been validated against a 4-socket Intel
Xeon X7460 Dunnington shared-memory machine.
An average absolute error of 25\% has been reported for the {\em SPLASH-2} benchmark suite (we report
18.77\%).

FastMP~\cite{FastMP}, is also an approximate simulator. It simulates a subset of cores in 
detail. Subsequently, it performs a real-time analysis of the behavior of the cores that have been
simulated in detail and uses this data to approximate
the behavior of the other cores. FastMP has been validated against a real x86 machine using the {\em SPEC 2006} 
benchmark suite. It suffers from an average error of 9.56\% but for some of the benchmarks the error is 
as high as 40\% (our maximum error is roughly 27\%).

To the best of our knowledge other popular simulators such as SESC~\cite{SESC} and MacSim ~\cite{MacSim} have not been validated.
Multi2sim, which is a heterogeneous simulator, has been validated for the GPU framework but
its CPU simulation framework has not been validated and published (to the best of our knowledge).

\section{Conclusion}
Tejas has been demonstrated to be a reliable simulator that provides a highly accurate
reflection of the simulated hardware. It must be noted that many details of the underlying
hardware are not known -- the branch predictor, the coherence protocol, the
NOC parameters, the select/data forwarding logic, to name a few. Better knowledge of these
will allow further reduction of the error in the accuracy.
The incomplete knowledge aside, Tejas shows an average absolute error of 11.45\% in serial benchmarks,
and 18.77\% in parallel benchmarks, which is 5-10 percentage points better than some of the most
popular architecture simulators currently in use (as of 2014).

\bibliographystyle{plain}
\bibliography{refs}
\end{document}